\documentclass[12pt]{article}
\pdfoutput=1

\usepackage{amsmath}
\usepackage{xcolor}
\usepackage{framed}
\definecolor{shadecolor}{rgb}{0.90,0.90,0.90}
\usepackage{cite}
\usepackage{subcaption}
\usepackage[utf8]{inputenc}
\usepackage[normalem]{ulem}
\usepackage[percent]{overpic}
\usepackage{mathtools}
\usepackage{setspace}
\usepackage{amsmath, amssymb, amsthm, float, graphicx}
\usepackage[linkbordercolor=blue]{hyperref}

\numberwithin{equation}{section}

\hypersetup{colorlinks=false, linkcolor=blue, citecolor=red}

\def\beq{\begin{eqnarray}}\def\eeq{\end{eqnarray}}
\def\be{\begin{equation}}\def\ee{\end{equation}}
\def\nn{\nonumber}

\def\b{\beta}
\def\d{\delta}

\def\D{\Delta}

\def\bz{\bar{z}}

\def\mc{{\mathcal{C}}}

\def\l{\eta_0}
\newcommand{\bmh}{\bar{\mathcal{H}}}
\newcommand{\mA}{\mathcal{A}}

\begin{document}
\begin{titlepage}
\title{\bf Four-point correlation modular bootstrap for OPE densities}
\date{}

\author{Carlos Cardona${}^{\Delta}$,  Cynthia Keeler${}^{\Delta}$,  William Munizzi${}^{\Delta}$\\~~~~\\
${}^{\Delta}$Physics Department, Arizona State University, \\Tempe AZ 85287 USA
}

\maketitle
\vskip 2cm
\abstract{In this work we apply the lightcone bootstrap to a four-point function of scalars in two-dimensional conformal field theory. We include the entire Virasoro symmetry and consider non-rational theories with a gap in the spectrum from the vacuum and no conserved currents. For those theories, we compute the large dimension limit ($h/c\gg1$) of the OPE spectral decomposition of the Virasoro vacuum.  We then propose a kernel ansatz that generalizes the spectral decomposition beyond $h/c\gg1$
. Finally, we estimate the corrections to the OPE spectral densities from the inclusion of the lightest operator in the spectrum.}

\vfill {\footnotesize 
	ccardon6@asu.edu,  keelerc@asu.edu,  wmunizzi@asu.edu}

\end{titlepage}
\newpage

\tableofcontents

\onehalfspacing

\newpage
\section{Introduction and motivation}

The study of conformal field theories in dimensions greater than two has seen a rapidly-increasing interest in recent years due to successful  application of the bootstrap program to conformally symmetric correlation functions,  revived in \cite{Rattazzi:2008pe, ElShowk:2012ht}.  In particular, by taking the lightcone limit of the crossing equation and expanding in large spin, \cite{Fitzpatrick:2014vua, Komargodski:2012ek, Alday:2016njk, Alday:2015ewa} obtained analytic results for the OPE coefficients and the anomalous dimensions of the spectrum. In this regime, the resultant quantities reveal properties universal to all conformal field theories in dimension three and higher. 

In two dimensions, the conformal symmetry extends to an infinite-dimensional Virasoro algebra. In the semi-classical limit, where the central charge is very large, this algebra reduces to the same finite-dimensional group $SO(d,1)$ as in the higher-dimensional case.  Thus, all the successful  techniques applied in higher dimensions can, in principle, be used in the semiclassical limit of any (non-rational) two-dimensional CFT.  However, applying these higher-dimensional techniques to the two-dimensional case does not produce a 
straightforward result due to several features of the two-dimensional theory. First, the two-dimensional theory has the infinite-dimensional Virasoro algebra instead of just the global conformal algebra. Next, in more than two dimensions,  all non-vacuum operators have higher weight.  In two dimensions,  instead,  all the operators in the Virasoro vacuum Verma module, such as the stress tensor and its composites, have zero weight; therefore a larger family of operators contributes in the same multiplet.

Two-dimensional CFTs themselves can be further divided into  rational and non-rational theories.  Techniques including integrability,  vertex algebras \cite{kac1998vertex} \footnote{For a nice recent account on the use of vertex algebras in this context see \cite{Cheng:2020srs}.},  and the quantum groups approach \cite{Ponsot:2000mt, Ponsot:1999uf} have led to impressive progress towards classifying the space of rational field theories.  Unfortunately for generic non-rational conformal field theories in two dimensions, this analysis is far from complete even in more-controllable regions of the parameter space, such as the lightcone limit. 

In order to translate the large spin results achieved in higher dimensions to the two-dimensional case,  we require a decomposition that accounts for the entire extended Virasoro symmetry.   A powerful tool that allows for a clean  analysis of large spin quantities in higher-dimensional conformal field theories is the Lorentzian inversion formula \cite{Caron-Huot:2017vep,  Simmons-Duffin:2017nub}, which shows the analyticity in spin of OPE coefficients (see \cite{Liu:2018jhs, Cardona:2018dov,  Cardona:2018qrt} 
 for earlier applications of the inversion formula to compute anomalous dimensions).  Unfortunately, we cannot directly apply this inversion formula to the two-dimensional case as it does not incorporate the entire Virasoro symmetry.  Remarkably,  progress is still possible:  there is  an analogous inversion formula relating t-channel to s-channel data that has been known for two decades \cite{Ponsot:2000mt, Ponsot:1999uf, Teschner:2001rv}.  By leveraging this impressive tool,  \cite{Collier:2018exn,  Collier:2019weq} studied the universal OPE large spin asymptotics for non-rational CFTs.
 
In the present work, we aim to reproduce some of these results via more traditional bootstrap methods.  Even though explicit forms for the Virasoro blocks are only known in particular limits, e.g. ~the semi-classical limit, their form is still constrained enough to justify certain conclusions regarding the large spin OPE spectral densities.  Previous work using the same strategy was done in \cite{Das:2017cnv,  Kusuki:2019avm, Kusuki:2018wcv },  but we are able to extend their results by proposing a kernel ansatz that goes beyond the strict semi-classical limit previously considered.  

Most of the previous work using the modular bootstrap focuses on the partition function; for example, \cite{Kraus:2016nwo,  Keller:2014xba} studied the asymptotic formula for the average value of light-heavy-heavy three-point coefficients,  in a generalization of Cardy’s formula for the high energy density of states, while \cite{Benjamin:2016fhe, Benjamin:2019stq,Cho:2017fzo,  Keller:2017iql,  Dyer:2017rul, Lin:2021udi} further generalized these results. For the four-point function, \cite{Das:2017cnv} follows a similar approach. Numerical work on the modular bootstrap has been undertaken in \cite{Collier:2016cls, Friedan:2013cba,  Hartman:2014oaa}, mostly focusing on the partition function. Our work here is highly motivated by these references.  

In this paper we study a slight modification of the lightcone limit, which we term the \emph{modular lightcone limit}.  In particular, we examine the crossing equation for the scalar four-point function, extracting the large spin OPE spectral densities using familiar bootstrap techniques that require the Virasoro conformal blocks.  We hope this paper will provide a step towards a generalization of the lightcone and the Euclidean bootstrap, both well understood in higher dimensions using conformal blocks, to the two-dimensional case where the Virasoro blocks are needed for computation. We begin by establishing notation and reviewing the Virasoro crossing equation in Section \ref{sec:Vir}.  In Section \ref{sec:Vac} we re-establish the results for the vacuum contribution to the OPE coefficient spectral density using a standard bootstrap approach, and in Section \ref{sec:BeyondVac} we extend these results to next order in the semi-classical limit by incorporating the first correction beyond the vacuum contribution.  We conclude and discuss the gravitational implications of our results in Section \ref{sec:Discussion}.

\section{Virasoro Crossing Equation}\label{sec:Vir}

We consider a scalar four-point function.  By conformal invariance,  this four-point function is given by
\be \label{4p}
\langle \prod_{i=1}^4{\cal O}_{\varphi}(x_i) \rangle = 
  \frac{1}{(x_{12}^2)^{\frac12(\Delta_1+\Delta_2)}(x_{34}^2)^{\frac12(\Delta_3+\Delta_4)}}
A(z,\bz),
\ee
where $A(z,\bz)$ characterizes the cross-ratio dependence.  The cross ratios themselves are given by
\be
z\bz={x_{12}^2x_{34}^2\over x_{13}^2x_{24}^2},\,\quad (1-z)(1-\bz)={x_{14}^2x_{23}^2\over x_{13}^2x_{24}^2}\,.
\ee

In two dimensions, fields factorize into holomorphic and anti-holomorphic components.  The total conformal dimension of an operator, $\D=h+\bar{h}$,  is the sum of its holomorphic  weight $h$ and its anti-holomorphic weight $\bar{h}$.  The operator's spin is $j=|h-\bar{h}|$, and  $c$ is the central charge of the CFT \footnote{Here we restrict to non-rational CFTs, therefore setting $c>1$ to avoid the minimal models.  When we do numerics, we further specialize to $c>25$. }. 

Here we study the OPE decomposition of the four-point correlation function  by a traditional bootstrap approach,  i.e.  by using Virasoro conformal blocks $\mathcal{F}$ and the crossing equation.  For the case where all external operators have the same conformal dimensions,  i.e. $\D_i=\D_0$ for  $i=1,\cdots, 4$, the crossing equation simplifies to
\begin{equation}
	\sum_{{\cal O}^s_i} C_{i}^2\, \mathcal{F}\left(h_i\middle|z\right)\bar{\mathcal{F}}\left(\bar{h}_j\middle|\bar{z}\right) = \sum_{{\cal O}^t_j} C_{j}^2\, \mathcal{F}\left(h_j\middle|1-z\right)\bar{\mathcal{F}}\left(\bar{h}_j\middle| 1-\bar{z}\right)\,.
\end{equation}
We introduce a spectral density of OPE coefficients defined as
\be\label{spectralC}
C(h)= \sum_{i}C_i\,\d (h-h_i),
\ee
where $h_i$ denotes the conformal dimension of the exchange operator ${\cal O}_i$, with an analogous definition for the barred coefficients.  In this way the crossing equation takes the form
\beq\label{crossing2}
	&&\int_{0}^{\infty}d h_s d\bar{h}_s \,C(h_s)^2\,\mathcal{F}\left(h_s\middle|z\right)\bar{\mathcal{F}}\left(\bar{h}_s\middle|\bar{z}\right)\nn\\
	 &&= 	\int_{0}^{\infty}d h_t d\bar{h}_t \,C(h_t)^2\, \mathcal{F}\left(h_t\middle|1-z\right)\bar{\mathcal{F}}\left(\bar{h}_t\middle| 1-\bar{z}\right).
\eeq

An important comment is in order here.   As will be demonstrated in subsequent sections,  we will obtain solutions to the spectral density which are approximated by smooth functions rather than linear combinations of distributions.  The proper statement would be that those solutions approximate the RHS of equation \eqref{spectralC} in a smeared sense,  i.e.  after integration against an appropriate test function.  The smearing mechanism required to make this statement rigorous has been recently explored through the use of Tauberian theorems, for example \cite{Mukhametzhanov:2019pzy,  Pal:2019zzr,  Das:2020uax}%
\footnote{ We give special thanks to Alex Maloney for providing a clear explanation of this issue.}.%

For the Virasoro blocks,  we will use the elliptic representation found by Zamolodchikov \cite{Zamolodchikov:426555}:
\be\label{Virasoros}
{\cal F}_{h_0}(h_s|z)=(1-z)^{\frac{c-1}{24}-2h_{0}}\left(z\right)^{\frac{c-1}{24}-2h_{0}}[\theta_{3}\left(q\right)]^{\frac{c-1}{2}-16h_{0}}\left(16q\right)^{h_s-\frac{c-1}{24}}H(h_s,q)\,,
\ee
where $q$  is known as the elliptic nome and $h_0$ is the conformal dimension of the external operators, while $h_s$ is the exchange dimension. The elliptic nome can be thought of as a conformal transformation:
\beq\label{nome}
q=e^{i\pi\tau(z)}\,, \qquad
\tau(z)=i{K(1-z)\over K(z)}\,,\qquad K(z)={1\over 2}\int_0^1{dt\over \sqrt{t(1-t)(1-zt)}}\,,
\eeq
so $K$ is an elliptic integral of the first kind.  In \eqref{Virasoros}, the function $\theta_{3}$ is the Jacobi theta function
\begin{equation}
    \theta_3(q) \equiv \sum_{n\in \mathcal{Z}}q^{n^2},
\end{equation}
and the function $H(h_s,q)$ in \eqref{Virasoros} is unknown in closed form, but can be computed recursively as a power expansion in $q$ to very high order \cite{Zamolodchikov:1987, Zamolodchikov:1985ie}.  In the semi-classical regime, where $h_s\gg c$, $H(h_s,q) \approx 1$; the overall prefactors  in \eqref{Virasoros} thus capture the semi-classical behavior.
Later we will also use the notation $\tilde{q}=q(-{1\over \tau})$,  corresponding to the modular S transformation.  

Given the elliptic representation for the Virasoro blocks \eqref{Virasoros},  we are mainly interested in two particular limits of the crossing equation.  For later convenience,  let us define  $\tau \equiv {i\beta\over \pi}$.   First we consider the limit $\b \to 0$, with $\bar{\b}$ fixed.   Then,  as we will show,  the limit $\bar{\b}\to 0$ (Euclidean) simply becomes a copy of $\b \to 0$, with all quantities replaced by their bar counterparts.  We then study the limit $\bar \b \to \infty$ while keeping  $\b\to 0$, which we term the modular lightcone limit due to its similarities with the better-known global lightcone limit.

\section{Spectral density OPE for vacuum}\label{sec:Vac}

In this section we study the OPE spectral density in the s-channel in the limit where the t-channel contribution is only due to the vacuum.  Specifically, we first take the $\b\to 0$ limit of  the four-point function crossing equation while leaving $\bar{\b}$ fixed, isolating the vacuum contribution in the t-channel; we will return to contributions beyond this vacuum limit in Section \ref{sec:BeyondVac}. In the $\b \to 0$ limit, we then study the s-channel contribution,  proposing  a kernel ansatze for the spectral density $C(h_s)$ in both the Euclidean limit $\bar \b \to 0$ and the modular lightcone limit $\bar \b \to \infty$. 
\subsection{Small $\b$ limit} \label{smallbeta}

In order to isolate the t-channel vacuum contribution, we need to consider the leading behavior of the conformal blocks when $\b\to 0$, as we now show. We begin by inserting the elliptic representation for the Virasoro blocks \eqref{Virasoros} into the crossing equation \eqref{crossing2}.  We note that the t-channel blocks depend on $(1-z)$, and thus they take the form \eqref{Virasoros} except with $z\leftrightarrow 1-z$ and $q\rightarrow \tilde q$, where $\tilde q = q(-1/\tau)$ as in \eqref{nome}.

To compare the terms independent of exchange dimension, we rewrite the $\theta_{3}\left(\tilde q \right)$ from the t-channel blocks using the modular transformation of the theta function
\be
 \theta_{3}\left(\tilde q\right)=\left({\beta\over \pi}\right)^{1/2}\theta_{3}\left(q\right),
\ee
where we have used $\tau={i\beta\over \pi}$ and $\tilde q = q(-1/\tau)$.
Accordingly, the terms independent of exchange dimension in \eqref{Virasoros} cancel when we plug into the crossing equation, up to powers of $\beta$ and $\bar\beta$.  Explicitly the crossing equation \eqref{crossing2} becomes
\beq\label{crossing1}
	&&\int_{0}^{\infty}d h_s d\bar{h}_s \,C(h_s)^2 \,\left(16\right)^{h_s+\bar{h}_s}\,
	e^{-\b(h_s-\frac{c-1}{24})}e^{-\bar{\b}(\bar{h}_s-\frac{c-1}{24})}H(h_s,e^{-\b})H(\bar{h}_s,e^{-\bar{\b}})\\
	 &&= \left(\frac{\b \bar \b}{\pi^2}\right)^{\frac{c-1}{4}- 8h_{0}}\int_{0}^{\infty}d h_t d\bar{h}_t \,C(h_t)^2 \left(16\right)^{h_t+\bar{h}_t}	e^{-{\pi^2}(h_t-\frac{c-1}{24})/\b}e^{-{\pi^2}(\bar{h}_t-\frac{c-1}{24})/ \bar{\b}}
	 H(h_t,e^{-\pi^2/\b})H(\bar{h}_t,e^{-\pi^2/\bar{\b}})\,.\nn
\eeq

In the limit $\b\to0$, the leading contribution in the holomorphic t-channel comes from the operator with smallest weight $h_t$, because any heavier operator is exponentially suppressed by the $e^{-1/\beta}$ terms. In general dimensions, the smallest weight operator would just be the vacuum; in two dimensions,  the vacuum Verma module captures contributions of all its descendants,  which includes the stress tensor.  We will nonetheless still use the term vacuum block to refer to it.


In order to have only vacuum block contributions at leading order in the small $\b$ limit of the t-channel, we also need to disallow contributions from representations with weight $h=0$ but $\bar h \neq 0$ (and vice versa). Stated in another way, we do not want to allow for conserved currents, because a CFT possessing a primary operator that is also a conserved current will have a vanishing gap between the vacuum block and the rest of the spectrum, as has been recently shown in \cite{Benjamin:2020swg}.       Instituting this restriction, the dominant contribution given by the  holomorphic vacuum block $h=0$ only couples to the corresponding anti-holomorphic vacuum block $\bar{h}=0$. The crossing equation thus simplifies to
\beq
	&&\int_{0}^{\infty}d h_s d\bar{h}_s \,C(h_s) C(\bar{h}_s)\left(16\right)^{h_s+\bar{h}_s}\,
	e^{-\b(h_s-\frac{c-1}{24})}e^{-\bar{\b}(\bar{h}_s-\frac{c-1}{24})}H(h_s,e^{-\b})H(\bar{h}_s,e^{-\bar{\b}})\nn\\
	 &&= \left(\b\bar \b /\pi^2\right)^{\frac{c-1}{4}-8h_{0}}e^{{\pi^2}(\frac{c-1}{24})/\b}e^{{\pi^2}(\frac{c-1}{24})/\bar\b}\,C(0)^2
	 H(0,e^{-\pi^2/\b})H(0,e^{-\pi^2 /\bar{\b}}).
\eeq
As we mentioned at the end of Section \ref{sec:Vir},  the function $H(h, q)$ can be computed as a series expansion in small $q$ recursively, and we have $H(h,q) \sim 1+\mathcal{O}(q)$. Since the limit $\b\to 0$ implies  $q\to 0$ in the t-channel blocks,  we can simplify the crossing equation even further, obtaining
\beq\label{crossing_total}
	&&\int_{0}^{\infty}d h_s d\bar{h}_s \,C(h_s)^2\,\left(16\right)^{h_s+\bar{h}_s}\,
	e^{-\b(h_s-\frac{c-1}{24})}e^{-\bar{\b}(\bar{h}_s-\frac{c-1}{24})}H(h_s,e^{-\b})H(\bar{h}_s,e^{-\bar{\b}})\nn\\
	 &&=(\b\bar \b/\pi^2)^{\frac{c-1}{4}-8h_{0}}e^{{\pi^2}(\frac{c-1}{24})/\b}e^{{\pi^2}(\frac{c-1}{24})/\bar\b}\,C(0)^2 H(0,e^{-\pi^2 /\bar{\b}}).
\eeq

\subsection{Saddle point}\label{holosaddle}
Just as in the lightcone limit in higher dimensions \cite{Komargodski:2012ek,  Fitzpatrick:2014vua},  we can see that the crossing equation \eqref{crossing_total} develops an essential singularity on the righthand side as $\b\to 0$.  On the lefthand side, all of the terms in the integrand can be expanded in small $\beta$.  The only way a Taylor series in $\beta$ can equal an essential singularity is if it has an infinite number of terms, so we should expect infinite contributions to the s-channel sum on the lefthand side.  Accordingly, this sum should be dominated by the tail, i.e. the s-channel expansion of the vacuum block must be dominated by large values of $h_s$.  We now verify this intuition with a saddle point analysis.

Before moving on to solving the crossing equation we have to address an important subtlety.  
In order to solve for the spectral densities,  we need an explicit form for the block $H(h_s,e^{-\b})$,  which is unfortunately not known in general.  However,  as we argued in the previous paragraph,  we mainly need its behavior for large values of $h_s$. Fortunately,  it turns out that the blocks simplify dramatically in this limit  \cite{Cardona:2020cfy}.  At leading order in inverse powers of  $h_s$,  the Virasoro blocks are approximated by%
\footnote{Beyond first order has been considered in \cite{Das:2020fhs}.}%
\be\label{virq}
H(h_s,\b)\equiv 1-{H_{-1}\over  h_s}\left({E_2(\b)-1\over 24}\right) +{\cal O}(1/h_s^2)\,,
\ee
where 
\be
H_{-1}={((c+1)-32 h_0) ((c+5)-32 h_0) \over 16}\,,
\ee 
and $E_2(\b)$ is an Eisenstein series of weight two. 

The needed limit $\b\to 0$ is not expected to commute with the large  $h_s$ approximation,  as has been recently argued numerically \cite{Das:2020uax}.  However, if we constrain ourselves to a region where the constant $H_{-1}$ in \eqref{virq} becomes small, i.e. either
$32 h_0\sim c+1$ or $32 h_0\sim c+5$, then the $\mathcal{O} (1/h_s)$ terms become small regardless of the value of $\b$.  If we constrain the relation between the external operator dimension $h_0$ and the central charge $c$ in this way, we can then take the limit $\b \to 0$ and then take $h_s$ large, at the cost of keeping $c$ finite.%
\footnote{See \cite{Das:2020uax} for details.}
Thus, at zeroth order in $1/h_s$,  with $h_0\sim {c+1\over 32 }$ and in the $\b\to 0$ limit,   the  crossing equation \eqref{crossing_total} becomes%
\footnote{The reader may be concerned that we have approximated our integrand at large $h_s$ without modifying the integral as a whole.  It turns out the saddle point expansion that follows will actually provide a justification for this approximation in hindsight. We can rewrite any integral $\int_0^{\infty} \mathcal{F}(\mathcal{H},e^{-\beta}) d\mathcal{H}$ as $\int_0^{\mathcal{H}_{\Lambda}} \mathcal{F}(\mathcal{H},e^{-\beta}) d\mathcal{H} + \int_{\mathcal{H}_{\Lambda}}^{\infty} \mathcal{F}(\mathcal{H},e^{-\beta}) d\mathcal{H}$, for some finite $\mathcal{H}_{\Lambda}$. $\int_0^{\mathcal{H}_{\Lambda}} \mathcal{F}(\mathcal{H},e^{-\beta}) d\mathcal{H} $ remains finite in the limit $\beta \rightarrow 0$; specifically it is bounded by the finite value $\int_0^{\mathcal{H}_{\Lambda}} \mathcal{F}(\mathcal{H},1) d\mathcal{H}$. For our case, $\int_{\mathcal{H}_{\Lambda}}^{\infty} \mathcal{F}(\mathcal{H},e^{-\beta}) d\mathcal{H}$ diverges as $\beta \rightarrow 0$, as long as we pick $\mathcal{H}_{\Lambda}$ below the saddle, as we can see from the divergence of the saddle itself.  Since an infinite contribution will always dominate over a finite one, the approximation within the integrand is justified.}%
\beq\label{smallb_largeh}
	&&\int_{0}^{\infty}d h_s \int_{0}^{\infty}d\bar{h}_s \,C(h_s)^2\, \left(16\right)^{h_s+\bar{h}_s}\,
e^{-\b(h_s-\frac{c-1}{24})}e^{-\bar{\b}(\bar{h}_s-\frac{c-1}{24})}H(\bar{h}_s,e^{-\bar{\b}})\nn\\
	 &&= (\b\bar \b/\pi^2)^{\frac{c-1}{4}-8h_{0}}e^{{\pi^2}(\frac{c-1}{24})/\b}e^{{\pi^2}(\frac{c-1}{24})/\bar\b}\,C(0)^2 H(0,e^{-\pi^2 /\bar{\b}})\, .
\eeq

We first examine the holomorphic part
\be
\int_{0}^{\infty}d h_s  \,C(h_s)\,\left(16\right)^{h_s}\,
e^{-\b(h_s-\frac{c-1}{24})}
= (\b/\pi)^{\frac{c-1}{4}-8h_{0}}e^{{\pi^2}(\frac{c-1}{24})/\b}\,C(0)\,.
\ee
Actually, we should be more precise; previous arguments  \cite{Collier:2016cls,  Benjamin:2019stq} have shown that any compact, unitary two-dimensional conformal field theory must have a gap smaller than $\mathcal{C}\equiv \frac{c-1}{24}$. Although it is a choice here, in the following section, the integration over $h_s$ `knows' about this maximal gap.  Accordingly, we will set the lower limit to be $\mathcal{C}$, not zero.  Rewriting in terms of $\mathcal{C}$, we have
\be\label{holomorphiccrossing}
\int_{\mathcal{C}}^{\infty}d h_s  \,C(h_s) \,\left(16\right)^{h_s}\,
e^{-\b(h_s-\mathcal{C})}
= (\pi/\b)^{\l}e^{{\pi^2}\mathcal{C}/\b}\,C(0)\,,
\ee
where we have additionally defined a shifted external operator weight $\l \equiv  8h_0 - \frac{c-1}{4}$ for future convenience.  The left hand side of this equation is now the Laplace transform of $C(h_s) 16^{h_s}$, with Laplace parameter $\beta$ and integration variable $h_s-\mathcal C$.  

Given the form of \eqref{holomorphiccrossing}, we can solve for the spectral OPE density $C(h)$ by applying the inverse Laplace transform. We find 
\be
C(h_s) 16^{h_s} = \frac{C(0)\pi^{\l}}{2\pi i} \int_{-i \infty}^{i\infty} \b^{-\l}\exp\left[\beta (h_s-\mathcal{C})+\frac{\pi^2 \mathcal{C}}{\b}\right]\, d\b.
\ee
The saddle point for this integral is given by
\begin{equation}\label{saddle1}
	\beta_s = \frac{\l}{(h_s - \mathcal{C})} \pm \frac{\sqrt{(-\l)^2 + 4\pi^2 \mathcal{C}(h_s - \mathcal{C})}}{2(h_s - \mathcal{C})}\sim  \frac{\l}{\mathcal{C}({h_s\over \mathcal{C}} - 1)} \pm \frac{\pi}{\sqrt{{h_s\over \mathcal{C}} - 1}}\,,
\end{equation}
%
We can see from this saddle point that the $\b\to 0$ limit is indeed dominated by large values of $h_s$.   As usual the integral at the saddle point approximation is a simple Gaussian that can be straightforwardly evaluated, giving the OPE spectral density in the large $h_s$ limit as
\be\label{spectral_saddle}
C(h_s) \, \sim {C(0)\over \left(16\right)^{h_s} \sqrt{\pi \mathcal{C}}}(\b_s)^{-\l-3/2}e^{\b_s(h_s-\mathcal{C})+ \frac{\pi^2}{\beta_s}\mathcal{C} }\,,
\ee
with $\b_s$ given by \eqref{saddle1}.  

In the limit  $h_s\gg\mathcal{C}$,  the second term from \eqref{saddle1} dominates. We can write the OPE spectral density in the semi-classical limit more explicitly as
\be\label{spectral_saddle_leading}
C(h_s) \, \sim {C(0)\over \left(16\right)^{h_s} \sqrt{\pi \mathcal{C}}}\left({h_s\over \mathcal{C}} - 1\right)^{-\l/2-3/4}e^{2\pi\mathcal{C}\sqrt{{h_s\over \mathcal{C}} - 1}}\,.
\ee
This result is in agreement with previous similar analyses done in \cite{Kraus:2016nwo,  Das:2017cnv, Kusuki:2018nms}.

\subsection{Kernel ansatz}
In section above,  we have computed an approximate solution for $C(h_s)$ by means of a saddle point analysis.  However,  we did need to assume that this saddle point method is valid for any sufficiently large value of $h_s$.  Examining the evaluated saddle point \eqref{saddle1},  we see the $\b\to 0$ limit is indeed reliable at this point when $h_s/\mathcal{C}\gg1$.  In this section, we propose a kernel ansatz that generalizes the result for $C(h_s)$ from the saddle point analysis and henceforth allows matching the t-channel vacuum via integration in the s-channel in a wider regime, down towards $h_s/\mathcal{C}\sim \mathcal{O}(1)$.  

We begin with the integral 
 \be\label{hologuess}
 \int_0^{\infty}d\mathcal{H}\,\cosh(2\pi  \sqrt{\mathcal{H}\mathcal{C}} )\,\mathcal{H}^{a^2}e^{- \b\mathcal{H}}
 =\beta ^{-a^2-1} \Gamma \left(a^2+1\right) \, _1F_1\left(a^2+1;\frac{1}{2};\frac{\mathcal{C} \pi ^2}{\beta }\right).
 \ee
The format of this integral matches the left hand side of the holomorphic crossing equation \eqref{holomorphiccrossing}, provided we 
introduce the convenient variable
\be
 \mathcal{H} \equiv h_s-\mathcal{C}\, .
\ee
We can match\footnote{The choice of integrand for the left hand side of \eqref{hologuess} is not unique, but rather belongs to a family of integrands that correctly reproduce the leading order terms on the right hand side after integration. When considering higher order terms, such as in \eqref{kernel_with_polynomial}, any appropriate selection of integrand yielding the correct matching order by order is sufficient for this calculation.} the right hand side of  \eqref{holomorphiccrossing} by taking the $\b \to 0$ limit of \eqref{hologuess}:
\be\label{hologuessLimit}
 \int_0^{\infty}d\mathcal{H}\,\cosh(2\pi  \sqrt{\mathcal{H}\mathcal{C}} )\,\mathcal{H}^{a^2}e^{- \b\mathcal{H}} 
 \approx {(\mathcal{C}\pi^2)^{a^2+1/2}\over \b^{2a^2+3/2}}e^{\pi^2\mathcal{C}\over \b}\,.
\ee
We set
\be
	a^2=\l/2-3/4=4 h_0-3 \mathcal{C}-3/4
\ee
by matching the $\b$ exponent in \eqref{holomorphiccrossing}.  We can then identify the spectral density reproducing the vacuum as
\begin{align}\label{kernel_vac}
C(h_s) &=  
 \frac{C(0) \pi^{1/2}}{16^{h_s} \mathcal{C}^{\eta_0/2 - 1/4}}\cosh\left[2\pi \sqrt{\mathcal{H}\mathcal{C}}\right](\mathcal{H})^{\eta_0/2-3/4}
\\
&=
\frac{C(0) \pi^{1/2}}{16^{h_s}\mathcal{C}^{1/2}}\cosh\left[2\pi \mathcal{C} \sqrt{\frac{h_s}{\mathcal{C}}-1}\right] \left(\frac{h_s}{\mathcal{C}}-1\right)^{\eta_0/2-3/4}.
\nn
\end{align}
As we anticipated in our saddle point analysis in \eqref{holomorphiccrossing},  the lower integration limit of \eqref{hologuess},    written in terms of $dh_s$, is $h_s= \mathcal{C}$.  Here we have a stronger justification: the integral `knows' about the maximal size of the gap via the kernel choice.  This lower bound is again in line with previous arguments about the size of the gap \cite{Collier:2016cls,  Benjamin:2019stq}. The vacuum in the t-channel is reproduced by the large $h_s$ tail of the integral, as expected from the saddle point analysis in Section \ref{holosaddle}. 

This result agrees with the saddle point analysis for $h_s\gg\mathcal{C}$. As we have explicitly shown, it also reproduces the vacuum block when $h_s\sim \mathcal{C}$.  Therefore, the kernel ansatz \eqref{kernel_vac} should be thought of as a generalization of the spectral density that is computed by an inverse Laplace transform followed by a saddle point analysis to fix the constants. 

As a final step,  we use the spectral densities to define an average value for the OPE coefficients \cite{Das:2017cnv}
\be\label{average0}
 \mathbb{C}(h_s) ={C(h_s)\over\mathcal{S}_0}\,,
\ee
where $\mathcal{S}_0$ is the asymptotic density of primary states.%
\footnote{For a more rigorous treatment of this average coefficients from Tauberian theorems of distributions, see \cite{Das:2020uax,  Mukhametzhanov:2019pzy} } %
 In other words, $\mathcal{S}_0$ is the fusion kernel for the vacuum character decomposition associated to the partition function  \cite{Zamolodchikov:2001ah,  Collier:2018exn}, given by
\be\label{partition_fusion}
\mathcal{S}_0=4\sqrt{2} \sinh\left(2\pi b \sqrt{\mathcal{H}}\right) \sinh\left(2\pi b^{-1}\sqrt{\mathcal{H}}\right),
\ee
where $b$ relates to the central charge as $c= 1+ 6(b+1/b)^2$.  This density of primaries is a refined version of Cardy's formula  \cite{Cardy:1986ie}.  In the limit $h_s/\mathcal{C}\gg1$ equation \eqref{average0} can be written
\be\label{average_asym}
 \mathbb{C}(\mathcal{H})\sim \mathcal{H}^{\l/2-3/4} e^{-2\pi \sqrt{\mathcal{HC}}}.
\ee
This result is compatible with previous results \cite{Das:2017cnv,  Collier:2018exn,  Collier:2019weq}. 
\subsection{Anti-holomorphic piece}
When evaluating the Euclidean limit of the anti-holomorphic piece, $\bar{\b}\to 0$, the analysis is exactly the same as was performed for its holomorphic counterpart. Therefore the result for the anti-holomorphic spectral density in the Euclidean limit is given by \eqref{kernel_vac} with $\bar{h}_s$ substituted for $h_s$.  The opposite limit,  when $\bar{\b}\to \infty$, is more interesting and more involved,  as we will demonstrate in this section. 

Consider the anti-holomorphic portion of our crossing equation \eqref{crossing_total}:
 \be
\int_{0}^{\infty}d\bar{h}_s \,C(\bar{h}_s)\left(16\right)^{\bar{h}_s}\,e^{-\bar{\b}(\bar{h}_s-\mathcal{C})}H(\bar{h}_s,e^{-\bar{\b}})= \left(\bar{\b}/\pi\right)^{6\mathcal{C}-8h_0}e^{\pi^2 \mathcal{C}/\bar\b} \,
\bar C(0)
H(0,e^{-\pi^2/\bar{\b}})\,.
\ee
As in the holomorphic case, we do not have a useful form for the Virasoro blocks at finite $q$ and finite $\bar{h}$.  However, taking the limit $\bar{\b}\to \infty$ we can approximate
 \be\label{AntiHoloCrossApprox}
\int_{0}^{\infty}d\bar{h}_s \,\bar{C}(\bar{h}_s)\left(16\right)^{\bar{h}_s}\,e^{-\bar{\b}(\bar{h}_s-\mathcal{C})}= (\bar{\b}/\pi)^{6\mathcal{C}-8h_0}e^{\pi^2\mathcal{C}/\bar\b} \,\bar{C}(0)H(0,\tilde{q}\to 1)\,.
\ee
Evaluating the last term in this limit is subtle, but we can estimate using numerics.  Restricting to the case of central charge $c=30$ and external dimensions $h_0=1$,  which is close to the point ${c+1\over 32}$,  we computed%
\footnote{We have computed up to order 50 for computational time convenience,  but we checked that above order $130$,  the expansion becomes asymptotic in the limit $\tilde{q}\to 1$.} %
 $H(0,\tilde{q})$ up to  $\tilde{q}^{50}$,  and have found that the following function provides a good fit for the Virasoro block:
\begin{figure}
\begin{center}
\includegraphics[scale=0.6]{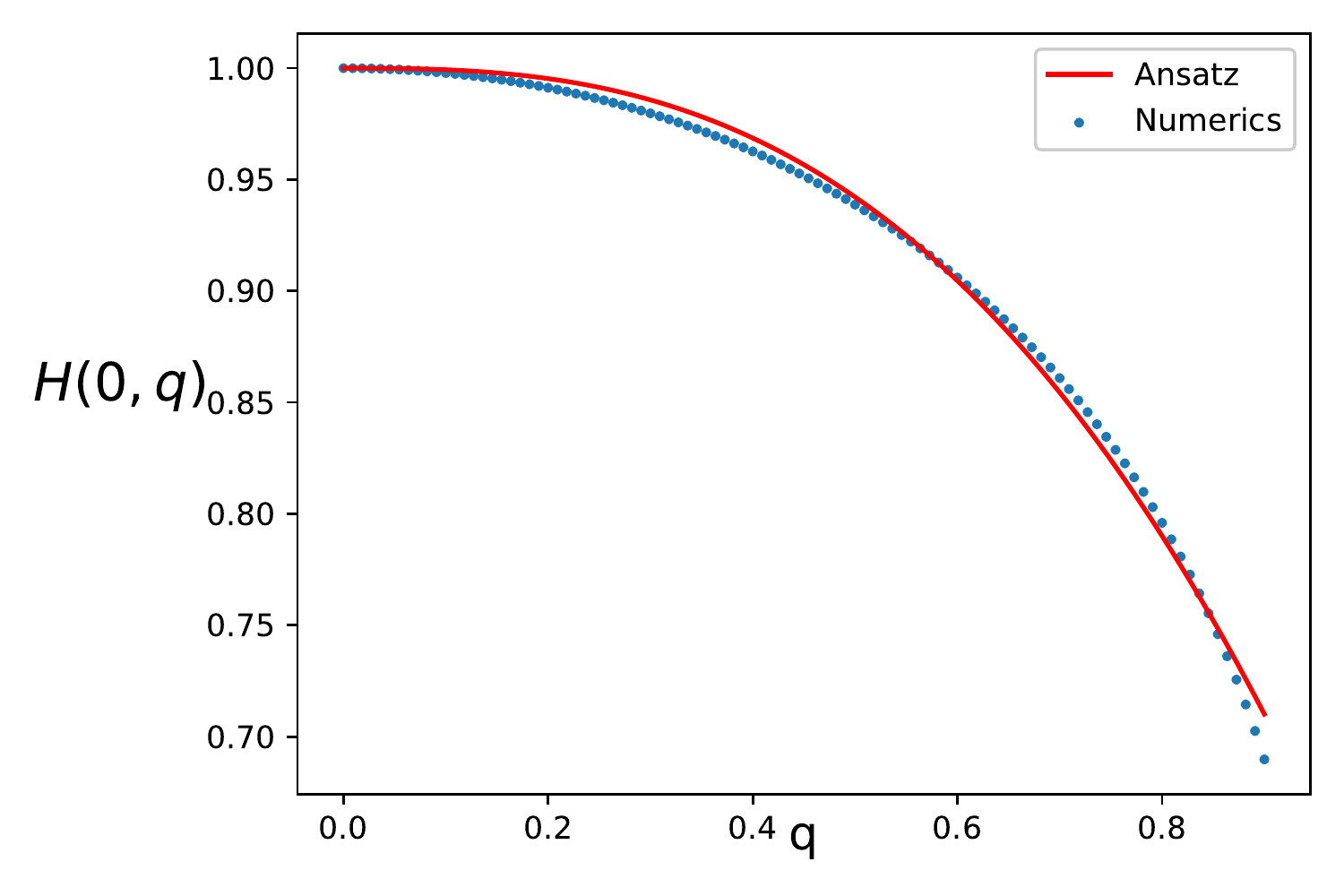}
\caption{ The dotted line corresponds to the vacuum Virasoro block computed numerically with the aid of Zamolodchikov recursion relations for the particular values $c=30$, $h_0=1$.  The solid red line is an approximation fit, explicitly given by equation \eqref{Hansatz}. }\label{H0vsq}
\end{center}
\end{figure}
\be\label{Hansatz}
H(0,\tilde{q})=1-a_1\,e^{-a_2\pi^2/\bar{\b}}\,.
\ee
Here we have introduced constants $a_1 = 0.8$ and $a_2 = 2.8$,  which are simply an approximation to the values given by the interpolating function fitting the numerical values for $H(0,\tilde{q})$.  If we instead did a numeric fit for the Virasoro block for different $c$ and $h_0$, we expect these values would change.  

%
\footnote{Here  we are not being too careful in finding the best fit to the numerical data,  since we just want to show how to deal with the anti-holomorphic piece once we interpolate the numerics.   We plan to  leverage the numerics in a more meaningful way in a future publication. } %
Inserting this approximation into the anti-holomorphic crossing equation \eqref{AntiHoloCrossApprox} we have 
 \be\label{cross_anti}
\int_{0}^{\infty}d\bar{h}_s \,\bar C(\bar{h}_s)\left(16\right)^{\bar{h}_s}\,e^{-\bar{\b}(\bar{h}_s-\mathcal{C})}= (\bar{\b}/\pi)^{6\mathcal{C}-8h_0}e^{\pi^2\mathcal{C}/\bar\b} \,\bar C(0)\left(1-a_1\,e^{-a_2\pi^2/\bar{\b}}\right)\,,
\ee
keeping in mind fixed parameters $h_0=1$ and $c=30$.  At leading order in $\bar{\b}\to \infty$ we can then equate
 \be\label{cross_anti_zero_order}
\int_{0}^{\infty}d\bar{h}_s \,\bar C(\bar{h}_s)\left(16\right)^{\bar{h}_s}\,e^{-\bar{\b}(\bar{h}_s-\mathcal{C})}= (\bar{\b}/\pi)^{6\mathcal{C}-8h_{0}}\,\bar C(0)\,.
\ee
We solve this equation using the same saddle point method as before,  employing a similar kernel ansatz as in the holomorphic analysis, with the important difference that we now take the  $\bar{\b}\to \infty$  limit of our result to compare with the expression above.  Explicitly, we consider the following integral
 \be\label{anitholoSimpleKernel}
\int_0^{\infty}d\bar{\mathcal{H}}\,\cosh(2\pi  \sqrt{\bar{\mathcal{H}}\mathcal{C}} )\,\bar{\mathcal{H}}^{\bar{a}^2}e^{- \bar{\b}\bar{\mathcal{H}}}= \bar{\b} ^{-\bar{a}^2-1} \Gamma \left(\bar{a}^2+1\right) \, _1F_1\left(\bar{a}^2+1;\frac{1}{2};\frac{\mathcal{C} \pi ^2}{ \bar{\b} }\right)\,,
\ee
where we have, as in the holomorphic sector, rewritten in terms of $\bar{\mathcal{H}}=\bar{h}_s-\mathcal{C}$, and set the lower bound at the maximal gap $\bar{h}_s=\mathcal{C}$.

In the limit $\bar \b \to \infty$, the argument of the hypergeometric function becomes $0$, so the lowest order term in $_1F_1$ is just one. Matching the lowest order power of $\bar \b$ with the right hand side of our crossing equation \eqref{cross_anti_zero_order},  we find 
\be\label{baraResult}
	\bar{a}^2=8 h_0-6\mathcal{C}-1=\l -1,
\ee
 where we have used our earlier definition of the shifted external weight $\l = 8h_0 - 6\mathcal{C}$. Solving for the OPE coefficient spectral density,  we find 
\begin{align}
\bar{C}(\bar{h}_s) &=\frac{\bar{C}(0)\pi^{\l}}{ 16^{\bar{h}_s} \Gamma\left(\l\right)}\bar{\mathcal{H}}^{\l-1}\cosh\left(2\pi  \sqrt{\bar{\mathcal{H}}\mathcal{C}} \right)
\nn\\
\label{kernel_vac_anti}&=  \frac{\bar{C}(0)\pi^{\l}}{ 16^{\bar{h}_s} \Gamma\left(\l\right)}\left(\bar{h}_s-\mathcal{C}\right)^{\l-1}\cosh\left(2\pi  \sqrt{(\bar{h}_s-\mathcal{C})\mathcal{C}} \right)\,.
\end{align}

This game can be continued to higher inverse-power orders in  $ \bar{\b}$. We wish to build a kernel ansatz for $\bar{C}({\bar{h}_s})$ that recovers the anti-holomorphic t-channel contribution on the righthand side of \eqref{cross_anti}. To simplify the algebra, we first expand this expression in powers of $x\equiv \pi^2\mathcal{C}/\bar{\beta}$: 
\begin{equation}
\label{Power_expansion_antiholo}
	\bar{C}(0) \left(\frac{\pi \mathcal{C}}{x} \right)^{-\l} \left(\sum_{k=0}^\infty \frac{x^k}{k!} \right) \left(1-a_1 \sum_{n=0}^\infty \frac{(-1)^n}{n!} \left(\frac{a_2 x}{\mathcal{C}} \right)^n \right).
\end{equation}

In order to match this t-channel expression, we need to upgrade the kernel \eqref{anitholoSimpleKernel}, to include a polynomial series in $\bmh$: 
\begin{align}
\label{kernel_with_polynomial}
	\int_0^\infty d\bmh \, \tilde{\mA} \cosh(2\pi \sqrt{\bmh\mc})& \sum_{n=0}^{\infty} \mA_n \bmh^{\bar{a}^2+n} e^{-\frac{\pi^2}{x}\mc\bmh}
	\\\nn
	&=\tilde{\mA} \left( \frac{\pi^2 \mc}{x}\right)^{-\bar{a}^2-1} \sum_{n=0}^\infty \mA_n \left(\frac{x}{\pi^2 \mc}\right)^n \Gamma(\bar{a}^2+1+n) \, {}_1F_1\left(\bar{a}^2+1+n;\frac{1}{2}; x \right)\, .
\end{align}
We will fix the constants $\tilde{\mA}$ and $\mA_n$ by expanding in powers of $x$ and matching the coefficients to \eqref{Power_expansion_antiholo}:
\begin{align}\label{crossing_with_sums}
\tilde{\mA} \left( \frac{\pi^2 \mc}{x}\right)^{-\l} \sum_{n=0}^\infty \mA_n \left(\frac{x}{\pi^2 \mc}\right)^n &
\sum_{k=0}^\infty\frac{ \Gamma(\l+n+k)\,\Gamma\left(\frac{1}{2}\right)}{\Gamma\left(\frac{1}{2}+k\right)}\frac{x^k}{k!}
\\\nn
&=\bar{C}(0) \left(\frac{\pi \mc}{x} \right)^{-\l} \left(\sum_{k=0}^\infty \frac{x^k}{k!} \right) \left(1-a_1 \sum_{n=0}^\infty \frac{(-1)^n}{n!} \left(\frac{a_2 x}{\mc} \right)^n \right).
\end{align}
Here we have already matched the lowest power of $x$, finding $\bar{a}^2+1=\l$ as in \eqref{baraResult}.
Matching the coefficient of the $x^{-\l}$ terms sets
\begin{equation}
	\tilde{\mathcal{A}} =\frac{\bar{C}(0) (1-a_1)}{\Gamma(\l)}\pi^{\l},
\end{equation}
where we have also chosen $\mA_0=1$.
Matching coefficients of the higher powers of $x$ in (\ref{crossing_with_sums}) gives a recursion relation for the coefficients $\mA_n$  for $n>0$:
\be
\mA_n=\frac{(\pi^2 \mc)^n \Gamma\left(\l\right)}{n! \,\Gamma \left(\l+n\right)}\left(\frac{1-a_1\left(1-\frac{a_2}{\mc}\right)^n}{1-a_1}\right)-\sum_{k=0}^{n-1} \frac{\left(\pi^2\mc\right)^{n-k}}{(n-k)!}\frac{\Gamma\left(\frac{1}{2}\right)}{\Gamma \left(\frac{1}{2}+n-k\right)}\mA_k\,.
\ee 
Solving this recursion relation for $\mathcal{A}_n$, we find the closed form solution
\be
\mA_n = \frac{\pi^{2n}\mc^n \Gamma(\l)}{1-a_1}\sum_{k=1}^n \left(\frac{1-a_1\left(1-\frac{a_2}{\mc}\right)^k}{k! \,\Gamma(\l +k)}-\frac{(1-a_1)\Gamma\left(\frac{1}{2}\right)}{k! \,\Gamma(\l) \,\Gamma\left(\frac{1}{2}+k\right)}\right)\sum_{\{\lambda_{n-k}\}}\prod_{i=1}^j \left(\frac{-\Gamma\left(\frac{1}{2}\right)}{\lambda_i ! \Gamma \left(\frac{1}{2}+\lambda_i\right)}\right)\,,
\ee
where the sum over $\left\{\lambda_{n-k}\right\}$ is taken over all sets of positive integers $\{\lambda_1, \lambda_2,\, \ldots ,\lambda_j\}$ such that $\sum_{i=1}^j \lambda_i=n-k$.  If $n-k=0$, this factor becomes 1.

Finally,  we can identify the anti-holomorphic OPE density as
\be
C(\bar{h}_s)= \tilde{\mA} \cosh\left(2\pi \sqrt{\bmh\mc}\right) \sum_{n=0}^{\infty} \mA_n \bmh^{\eta_0+n-1}\,.
\ee
Even though this result depends on the numerical fitting \eqref{Hansatz} with particular values $a_1=0.8,\,a_2=2.8$,  we have observed that for a wide enough range of values $\{h_0, c\}$,  the numerical  anti-holomorphic vacuum block can be approximated by the same function  \eqref{Hansatz}  at different values of the fitting constants $a_1$ and $a_2$.  The fact that we have found a closed form expression for the coefficients $\mA_n$, as functions of $a_1$ and $a_2$,  indicates a fruitful path towards numerically exploring the OPE density vacuum contribution in the modular lightcone limit.

\section{Spectral density OPE beyond vacuum contribution}\label{sec:BeyondVac}

So far we have considered OPE spectral densities that reproduce the vacuum block in the crossed channel, provided a gap separates this block from the rest of the spectrum.  We now go beyond the vacuum contribution in the t-channel, allowing a contribution from the Virasoro block associated to the primary operator closest to the vacuum.  We refer to this next-heaviest block as the lightest operator block, and we now compute the correction to the spectral density arising (in the s-channel) when we consider this lightest operator block in the t-channel of the crossing equation.

We first use a saddle point approximation to solve for the leading order correction for the spectral densities at $h_s/\mathcal{C}\gg1$ due to the presence of the  lightest operator block.  We then propose a generalization beyond $h_s/\mathcal{C}\gg1$ by introducing a kernel ansatz for the densities.
\subsection{Saddle point}\label{NonVacuumSaddleSection}
We start the discussion by writing the crossing equation explicitly in this case.  In Section \ref{smallbeta}, the $\b\to0$ limit, at leading order in small $\b$ and large exchange dimension  \eqref{smallb_largeh}, gave only the vacuum contribution in the t-channel.  If we want to include the next order corrections to the crossing equation \eqref{crossing_total} in small $\beta$,  a similar argument leads us to%
\footnote{In this section we are going to consider the holomorphic sector only,   mainly due to the fact that for the anti-holomorphic correction we don't have much information and  would be forced to rely on numerics alone. }%
\beq\label{hol_crossing_correction}
&&\int_{0}^{\infty}d h_s  \,\left(C^{0}(h_s)+\delta C^{0}(h_s)\right)\, \left(16\right)^{h_s}\,
e^{-\b(h_s-\frac{c-1}{24})}\nn\\
&&= (\b/\pi)^{\frac{c-1}{4}-8h_{0}}e^{{\pi^2}(\frac{c-1}{24})/\b}\,C(0) 
+\,(\b/\pi)^{\frac{c-1}{4}-8h_{0}}16^{h_{min}}C(h_{min})	e^{-{\pi^2\over\b}(h_{min}-\frac{c-1}{24})}.
\eeq
The new term $\delta C^0 (h_s)$ in the s-channel represents the corrections to the spectral OPE density produced from the lightest operator in the t-channel,  whose block contribution is given by
the second term in the second line.%
\footnote{ In the second line of \eqref{hol_crossing_correction} we have neglected higher order terms in small $\b$ from expanding $H\sim 1+\mathcal{O}(q)$, both when multiplied by the vacuum and the lightest operator.   It is possible that the $\mathcal{O}(q)$ terms in $C(0)\, H$ dominate over the leading term $C(h_{min})$ for the lightest operator.  However, here and in previous sections we have considered $h_{min}<{\cal C}$  with ${\cal C}$ to be of order one or less, while the $\mathcal{O}(q)$ term in $H$ has a more negative exponent set by a number greater than one, e.g. $a_2$ in \eqref{Hansatz}.  Thus, the second term in the second line of \eqref{hol_crossing_correction} is the dominant correction.  } %
The vacuum is of course already solved by \eqref{kernel_vac},  or in other words,  the first integration term at the first line of \eqref{hol_crossing_correction} equals the first term in the second line and therefore the correction terms satisfy
\be\label{crossing_beyond_vac}
\int_{0}^{\infty}d h_s  \,\delta C(h_s) \left(16\right)^{h_s}\,
e^{-\b(h_s-\frac{c-1}{24})}=(\b/\pi)^{\frac{c-1}{4}-8h_{0}}16^{h_{min}}C(h_{min})	e^{-{\pi^2\over\b}(h_{min}-\frac{c-1}{24})}\,.
\ee
By using again an inverse Laplace transform to solve for $\delta C$,   we find the saddle point of the resulting integral over $\b$ to be at
\be 
\b_s=\frac{4
  h_0-3 \mathcal{C}}{(h_s-\mathcal{C})}+\frac{\sqrt{(8 h_0-6 \mathcal{C})^2-4\pi ^2 (h_s-\mathcal{C}) \left(  h_{min}-\mathcal{C}\right)}}{2 (h_s-\mathcal{C})}\sim\frac{4h_0-3 \mathcal{C}}{(h_s-\mathcal{C})}+\pi\sqrt{  ( \mathcal{C}-h_{min})\over (h_s-\mathcal{C})}\, .
\ee
Notice that here we need $h_{min}<\mathcal{C}$ for the saddle to be real,  which is nevertheless automatically satisfied by the minimal gap.

The saddle point computation  leads to,
\be 
\d C(h_s) \sim {C(h_{min})16^{h_{min}-h_s }\over  \sqrt{\pi(\mathcal{C}-h_{min})}}\b_s^{6\mathcal{C}-8 h_0+3/2}\,e^{\b_s(h_s-\mathcal{C})-{\pi^2\over \b_s}(h_{min}-\mathcal{C})}
\ee
Taking the dominant second term in the saddle point, we can write this as
\be\label{saddle_beyond_vac}
\d C(h_s) \sim {C(h_{min})16^{h_{min}}\over  \sqrt{\pi(\mathcal{C}-h_{min})}}\left(\sqrt{  ( \mathcal{C}-h_{min})\over (\mathcal{C}-h_s)}\right)^{6\mathcal{C}-8 h_0+3/2}\,e^{2\pi\sqrt{  ( \mathcal{C}-h_{min})(h_s-\mathcal{C})}}\,.
\ee
As in the vacuum case,  the saddle point result is reliable as long as $h_s/\mathcal{C}\gg1$.  We would like to  generalize the spectral density OPE correction using the same technique as for the vacuum block contribution by proposing a similar spectral density ansatz.
\subsection{Kernel  ansatz}
Based on the saddle point analysis of Section \ref{NonVacuumSaddleSection} and the results from \cite{Collier:2019weq,  Collier:2018exn},  we propose the following ansatz as a  generalization for the correction from the lightest operator to the spectral density:
\be 
\d C(h_s) \sim C(h_{min})16^{h_{min}-h_s }\left((h_{min}- \mathcal{C})^{3\mathcal{C}-4 h_0+1/4}\over (\mathcal{C}-h_s)^{3\mathcal{C}-4 h_0+3/4}\right)\,
\cosh\left({2\pi\sqrt{ h_s-\mathcal{C}}}\right)^{\sqrt{  \mathcal{C}-h_{min}}}\, .
\ee
This ansatz reproduces the saddle point result  \eqref{saddle_beyond_vac} in the limit ${h_s\over \mathcal{C}}\gg1$.  However,  a stronger  check would be to prove that the integration over $h_s$ reproduces the lightest $h_{min}$ block in the t-channel.   Specifically, we want to perform the integral
\be\label{kernel_int_bv}
{C(h_{min})16^{h_{min} }\over (h_{min}- \mathcal{C})^{-3\mathcal{C}+4 h_0-1/4}}\int_{0}^{\infty} \,dh_s\,\left(1\over (\mathcal{C}-h_s)^{3\mathcal{C}-4 h_0+3/4}\right)\,
\cosh\left({2\pi\sqrt{ h_s-\mathcal{C}}}\right)^{\sqrt{  \mathcal{C}-h_{min}}}e^{-\b (h_s-\mathcal{C})}\,.
\ee
\begin{figure}
	\centering
		\begin{overpic}[scale=0.8]{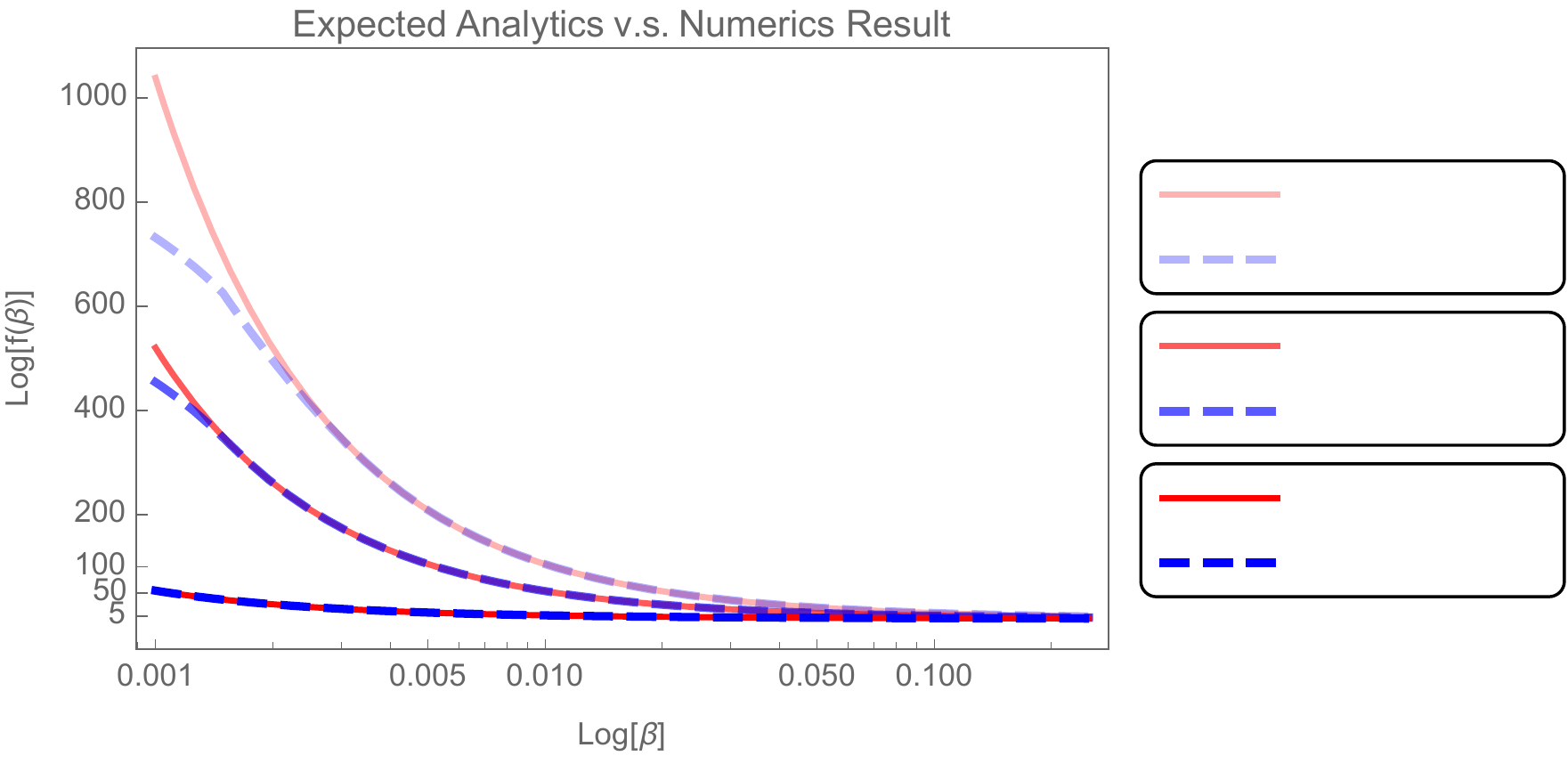}
			\put (82,33.45) {$h_{min} = 0.8 \mathcal{C}$}
			\put (82,23.75) {$h_{min} = 0.9 \mathcal{C}$}
			\put (82,14) {$h_{min} = 0.99 \mathcal{C}$}
		\end{overpic}
		\caption{Comparison of analytic t-channel crossing with results of numerical integration over kernel ansatz for relevant values of $h_{min}$. The solid red lines display the functional form expected from the t-channel crossing, given by the righthand side of \eqref{crossing_beyond_vac}. The dashed blue lines exhibit successive results of numerically integrating over the kernel \eqref{kernel_int_bv}. }
		\label{fig:NumericsFittingWithHmin}
\end{figure}
While we did not find a clever way to perform this integration, we have numerically evaluated the integral for several values of the difference $h_{min} - \mathcal{C}$ in the range $(0,1)$.  In figure \ref{fig:NumericsFittingWithHmin} we display three such cases for central charge $c=30$ and external dimension $h_0 =1$,  observing convincing agreement between the numerical result and the analytic t-channel of the crossing equation \eqref{crossing_beyond_vac}.  As the value of $h_{min}$ approaches the value of $\mathcal{C}$, the analytic expectation becomes closer to the numerical integration. This behavior is expected, since the smaller  $\mathcal{C}-h_{min}$ becomes,  the less suppressed is the contribution from the lightest operator;  that is,  for smaller $\mathcal{C}-h_{min}$, the ansatz is a better approximation in the range of smaller  $\b'$s,  which is our initial limiting condition.  When the difference between $\mathcal{C}$ and $h_{min}$ is increased,  deviation between the plots occurs for larger values of $\beta$,  which is evident from the saddle point  \eqref{saddle_beyond_vac}.

Beyond the numerical check,  we can, one last time, resort to a saddle point analysis. This time the saddle point analysis is on the integration \eqref{kernel_int_bv} over $h_s$ in the asymptotic limit, so we find 

\beq
I(\b)&=&\int_{0}^{\infty} \,dh_s\,\left( \mathcal{C}-h_s\right)^{-3\mathcal{C}+4 h_0-3/4}\,
e^{{2\pi\sqrt{ (h_s-\mathcal{C})(\mathcal{C}-h_{min})}}}e^{-\b (h_s-\mathcal{C})}\nn\\
&\sim &{2\pi(\mathcal{C}-h_{min})^{1/2}\over \b^{3/2} } e^{\pi^2\left({\mathcal{C}-h_{min}\over \b}\right)} \left({\pi^2(h_{min}-\mathcal{C})\over\b^2}\right)^{-3\mathcal{C}+4 h_0-3/4}\nn\\
&=& e^{\pi^2\left({\mathcal{C}-h_{min}\over \b}\right)} \left({(h_{min}-\mathcal{C})^{-3\mathcal{C}+4 h_0-1/4}\over\b^{(-6\mathcal{C}+8h_0)}}\right)\,.
\eeq
From this result, we obtain the right hand side of  \eqref{crossing_beyond_vac} after multiplying by the overall factor in front of the integral \eqref{kernel_int_bv}.  

Just as in the definition \eqref{average0},  we now define an average correction by dividing by the generalized Cardy formula \eqref{partition_fusion},  namely,
\be 
\d \mathbb{C}(h_s)\equiv{\d C(h_s)\over \mathcal{S}_0}\,.
\ee
Taking the limit ${h_s\over \mathcal{C}}\gg1$, 
\be
\mathcal{S}_0=4\sqrt{2} \sinh\left(2\pi b \sqrt{(h_s-\mathcal{C})}\right) \sinh\left(2\pi b^{-1}\sqrt{(h_s-\mathcal{C})}\right) \to e^{\left(4\pi \sqrt{\mathcal{C}(h_s-\mathcal{C})}\right) },
\ee
so we find 
\be
\d \mathbb{C}(h_s) \sim {C(h_{min})16^{h_{min}-h_s }\over  \sqrt{\pi(\mathcal{C}-h_{min})}}\left(\sqrt{  ( \mathcal{C}-h_{min})\over (\mathcal{C}-h_s)}\right)^{6\mathcal{C}-8 h_0+3/2}\ e^{-4\pi\sqrt{\mathcal{C}(h_s-\mathcal{C})}\left(1-{1\over 2}\sqrt{1-{h_{min}\over \mathcal{C}}}\right) }.
\ee
This result supports the claim that corrections to the spectral densities from including  non-vacuum contributions  are suppressed with respect to the leading contribution from the vacuum \eqref{spectral_saddle_leading}.  In fact, these corrections are exponentially suppressed.  This result corresponds to the analogous result for the one point function  in \cite{Kraus:2016nwo}. 
\section{Discussion and conclusions}\label{sec:Discussion}

In this paper we studied OPE spectral densities for the four-point correlation function of scalars in the large exchange dimension limit.  Our technique was to solve the modular bootstrap in an appropriate limit (the so-called modular lightcone limit),  allowing us to decouple the Virasoro vacuum  from the rest of the conformal dimensions spectrum.  We further restricted ourselves to theories that do not have conserved currents, and insisted on a twist gap in the spectrum,  allowing us to identify 
the contributions to the spectral densities of OPE coefficients at large spin from the Virasoro vacuum.

First we solved the crossing equations by resorting to a saddle point computation that allowed us to capture the OPE spectral density in the limit where the  dimension of the operator being exchanged is large in units of ${\cal C}$,  i.e.  ${h\over \mathcal{C}}\gg1$.  We then use this result as a leverage point to generalize the spectral densities beyond that region and towards ${h\over \mathcal{C}}\sim1$ by proposing an ansatz  that solves the crossing equations in the modular lightcone limit.   Just as the Cardy formula measures the density of primary operators in the large dimension limit,  our results can be understood as an extension of the Cardy formula to the density of OPE primary coefficients.

An obstruction to the development of a full-fledged large spin  perturbation theory comes from the lack of a practical closed form for the Virasoro blocks.  However,  we have shown in this paper that despite the limited knowledge we have of the blocks,  it is sufficient to allow for a leading order analysis.  We were able to perform an analytical analysis in the holomorphic sector of the OPE expansion,  but needed some numerical aid for the counterpart in the anti-holomorphic sector of the lightcone limit.  The numerical data needed for the Virasoro blocks was been obtained by solving the Zamolodchikov recursion relations numerically.

Recently,  some results for the Virasoro blocks at large exchange dimension have surfaced \cite{Cardona:2020cfy,  Das:2020fhs, Das:2020uax, Kashani-Poor:2013oza},  which offer hope of going beyond leading order in the large spin analysis, at least numerically.  We have in fact already used some of these results in the main body of the text.

Importantly,  we did not make any further assumptions on the theory under consideration,  beyond the existence of a gap separating the Virasoro vacuum from the rest of the spectrum and the absence of global conserved currents.  Henceforth,  we can think of our results as universal up to those assumptions.  Recently it was shown that this universality for the OPE coefficients of heavy operators is nicely captured by the DOZZ OPE of Liouville theory \cite{Collier:2019weq}.  In particular,  our result \eqref{average_asym} can be written in terms of the DOZZ coefficient.

Although we do not directly explore a gravitational interpretation of our work,  we expect it to provide a reliable semi-classical description of $AdS_3$ gravity even at finite values of $c$ \cite{Kraus:2016nwo,  Ghosh:2019rcj}.  Along the same lines,  it would be interesting to derive some of the results of this paper from a purely gravitational analysis,  in particular through the computation of the Witten diagrams corresponding to the four-point function of scalars considered here,  by using several of the methods developed recently  \cite{Costantino:2020vdu,Yuan:2017vgp,Carmi:2019ocp,Cardona:2017tsw,  Bertan:2018khc}.  Our results will be useful for studies of coarse-graining CFTs to produce gravitational duals, along the lines of \cite{Kraus:2016nwo,Das:2017cnv,Benjamin:2019stq,Miyaji:2021ktr, Benjamin:2021wzr}.  In those papers,  a coarse-grained average of the CFT result is compared to a black hole geometric result.  We specifically think it would be interesting to consider if a set of (non-minimal, as we consider) CFTs has a condition for successful coarse graining to a gravity theory,  where  the condition of $c=c_{crit}$ would be set by the asymptotic behavior of three-point spectral density instead of  the asymptotic density of states,  as proposed in \cite{Benjamin:2021wzr}.

\newpage
\section*{Acknowledgments}
We thank Alex Maloney and Eric Perlmutter for useful conversations.   This work was supported by the U.S. Department of Energy, Office of High Energy Physics, under Award No. DE-SC0019470 and C.C. by the National Science Foundation Award  No. PHY-2012195 at Arizona State University.

\bibliographystyle{JHEP}
\bibliography{VirasoroBib}
\end{document}